\documentclass[graybox]{svmult}

\usepackage{graphicx}        % standard LaTeX graphics tool
\usepackage{multicol}        % used for the two-column index
\usepackage[bottom]{footmisc}% places footnotes at page bottom

\usepackage{cite,url}

\usepackage{amssymb}
\usepackage{amsmath}
\usepackage{amsfonts}
\usepackage{arydshln}
\usepackage{epstopdf}

\newcommand{\be}{\begin{equation}}
\newcommand{\ee}{\end{equation}}
\newcommand{\bg}{\begin{svgraybox}}
\newcommand{\eg}{\end{svgraybox}}
\newcommand{\bd}{\begin{displaymath}}
\newcommand{\ed}{\end{displaymath}}

\newcommand{\Hi}{{\cal H}_\infty}

\newcommand{\beqa}{\begin{eqnarray}}
\newcommand{\eeqa}{\end{eqnarray}}
\newcommand{\bena}{\begin{eqnarray*}}
\newcommand{\eena}{\end{eqnarray*}}
\newcommand{\beqn}{\begin{equation} \label}
\newcommand{\eeqn}{\end{equation}}
\newcommand{\bap}{\left[ \begin{array}}
\newcommand{\eap}{\end{array} \right]}
\newcommand{\mato}{\left( \begin{array}}
\newcommand{\matc}{\end{array} \right)}
\newcommand{\BLOCK}[4]{\left(\begin{array}{cc} #1 & #2 \\ #3 & #4 \end{array}\right)}
\newcommand{\bea}{\begin{array}}
\newcommand{\eea}{\end{array}}

\newcommand{\smallmato}{\left( \begin{smallmatrix}}
\newcommand{\smallmatc}{\end{smallmatrix} \right)}

\newcommand{\LL}{{\cal L}}

\newcommand{\inv}{^{-1}}
\newcommand{\diag}{{\rm Diag}}

\begin{document}

\title*{Computer Aided Control System Design for Time Delay Systems using MATLAB\textsuperscript{\textregistered}}
 \titlerunning{CACSD for Time Delay Systems using MATLAB}

\author{Suat Gumussoy and Pascal Gahinet}
\institute{Suat Gumussoy and Pascal Gahinet \at MathWorks, \email{{suat.gumussoy, pascal.gahinet}@mathworks.com}
\and
This work is based on and the extension of the Authors' conference paper \cite{GumussoyIFACTDS2012}.
}

\maketitle

\abstract{Computer Aided Control System Design (CACSD) allows to analyze complex interconnected systems and design controllers achieving challenging control requirements. We extend CACSD to systems with time delays and illustrate the functionality of Control System Toolbox in MATLAB for such systems. We easily define systems in time and frequency domain system representations and build the overall complex system by interconnecting subsystems. We analyze the overall system in time and frequency domains and design PID controllers satisfying design requirements. Various visualization tools are used for analysis and design verification. Our goal is to introduce these functionalities to researchers and engineers and to discuss the open directions in computer algorithms for control system design.}

\section{Introduction}
Time delays are frequently seen in many control applications such as process control, communication networks, automotive and aerospace, \cite{Dugard1998,NiculescuBook,ErneuxBook}. Depending on the delay length, they may limit or degrade the performance of control systems unless they are considered in the design, \cite{GuBook,ComplexTDSBook}. Although  considerable research effort is devoted to extend classical and modern control techniques to accommodate delays, most available software packages for delay differential equations (DDE) \cite{retard,enright,BIFTOOLManual20,BredaTraceDDE09} are restrictive and not developed for control design purposes.

We present the currently implemented framework and available functionality in MATLAB for computer-aided manipulation of linear time-invariant (LTI) models with delays. We illustrate this functionality for  each important step in every practical control design:
\begin{itemize}
\item system representations in time and frequency domains,
\item interconnections of complex systems,
\item analysis tools and design techniques for time delay systems.
\end{itemize}
By introducing available functionality in Control System Toolbox, our goal is to facilitate the design of control systems with delays for researchers and engineers. Moreover, we discuss possible enhancements in CASCD for time delay systems, to illustrate the gap between the desired analysis / design techniques and the current control software implementation.

At the heart of this framework is a linear fractional transformation (LFT) based representation of time delay systems \cite{lft}. This representation handles delays in feedback loops and is general enough for most
control applications. In addition most classical software tools for analyzing delay-free LTI systems
are extended to this class of LTI systems with delays. Given the widespread use of linear techniques in control system design, this framework and the accompanying software tools should facilitate CACSD in the presence of delays, as well as stimulate more research into efficient numerical algorithms for assessing the properties and performance of such systems.

\section{Motivation Examples} \label{sg:sec:motivation}
\begin{figure}[h]
\begin{center}
\resizebox{8.0cm}{!}{\includegraphics{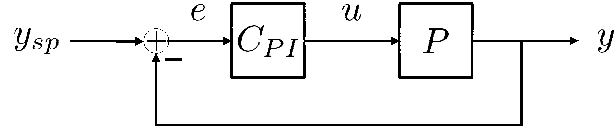}}
\caption{Feedback loop with PI controller.\label{sg:fig:cs1}}
\end{center}
\end{figure}

A standard PI control example is given in \cite{tankexample} where the plant is a chemical tank and a single-input-single-output system with an input-output delay (i.e., dead-time system),
\be \label{sg:theplant}
P(s) = e^{- 93.9 s} {5.6 \over 40.2 s+1}.
\ee In the classical feedback configuration in Figure~\ref{sg:fig:cs1}, the standard PI controller is chosen as
\be \label{sg:PI1}
C_{PI}(s) = K (1 + {1 \over T_i s})
\ee where $K=0.1$ and $T_i=100$. The closed-loop transfer function from $y_{sp}$ to $y$ is
\[
T_{PI}(s)=\frac{4020s^2+100s}{4020s^2+100s+(56s+0.56)e^{-93.9s}}.
\] This transfer function has an \emph{internal} delay which can not be represented by input or output delays. Therefore, the representation for time delay systems has to capture this type of systems and to be \emph{closed} under block diagram of operations.
This plant and the controller in MATLAB are defined as
\bg
\vspace{-.4cm}
{\small
\begin{verbatim}
  P = tf(5.6,[40.2 1],'OutputDelay',93.9);   % plant
  Cpi = 0.1 * (1 + tf(1,[100 0]));           % PI controller
\end{verbatim}
}
\vspace{-.2cm}
\eg \noindent and the closed-loop system $T_{PI}$ is obtained by the {\tt feedback} command:
\bg 
\vspace{-.4cm}
{\small
\begin{verbatim}
  Tpi = feedback(P*Cpi,1);      % Closed-loop transfer, ysp -> y
\end{verbatim}} 
\vspace{-.2cm}
\eg Note that these commands are natural extensions of delay-free case and are used for systems with delays without new syntax for the user.

The MIMO time delay systems may have different transport delays for each input-output channel, i.e.,
\[
H(s)=\left(
       \begin{array}{cc}
         e^{-0.1s}\frac{2}{s} & e^{-0.3s}\frac{s+1}{s+10} \\
         10 & e^{-0.2s}\frac{s-1}{s+5} \\
       \end{array}
     \right).
\] We define such systems in MATLAB by the following commands:
\bg
\vspace{-.4cm}
{\small
\begin{verbatim}
s = tf('s');
H = [2/s (s+1)/(s+10); 10 (s-1)/(s+5)];   % delay-free system
H.ioDelay = [0.1 0.3; 0 0.2];             % transport delays
\end{verbatim}}
\vspace{-.2cm}
\eg

We see on our motivation examples that the representation of time delay systems has certain challenges. Next section, we present the LFT-based representation of time delay systems to address these challenges and discuss its advantages.

\section{System Representation} \label{sg:sec:representation}
We represent time delay systems by the linear-fractional transformation (LFT). Recall
that the LFT is defined for matrices by
\[
\LL(\BLOCK{M_{11}}{M_{12}}{M_{21}}{M_{22}},\Theta) :=
M_{11} + M_{12} \Theta (I-M_{22}\Theta)\inv M_{21} \; .
\]
The LFT has been extensively used in robust control theory for representing models
with uncertainty, see \cite{lft} for details.

\begin{figure}
\begin{center}
\resizebox{8.0cm}{!}{\includegraphics{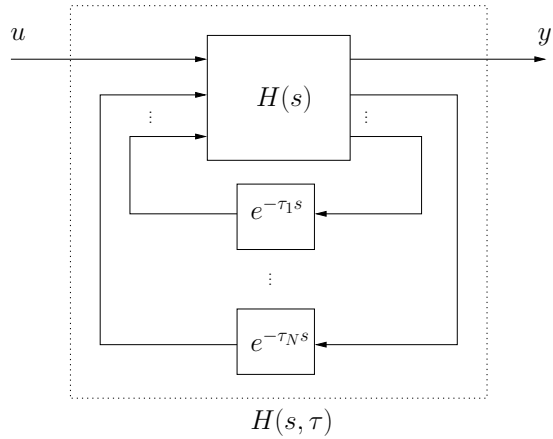}}
\caption{LFT-based modeling of LTI systems with delays.\label{sg:fig1}}
\end{center}
\end{figure}

Consider the class \emph{generalized LTI} (GLTI) of continuous-time LTI systems whose transfer function is of the form
\beqa
H(s,\tau) = \LL ( \underbrace{\BLOCK{H_{11}(s)}{H_{12}(s)}{H_{21}(s)}{H_{22}(s)}}_{H(s)} ,
\Theta(s,\tau) )
\nonumber
\\
\hspace*{.5cm}
\Theta(s,\tau) := \diag \left( e^{-\tau_1 s} , \ldots , e^{-\tau_N s} \right)
\label{sg:genlti_lft}
\eeqa
where $H(s)$ is a rational (delay free) MIMO
transfer function, and $\tau = ( \tau_1,\ldots,\tau_N )$ is a
vector of nonnegative time delays. Systems in this class are modeled as the LFT interconnection of a delay-free
LTI model and a bank of pure delays (see Figure \ref{sg:fig1}). As such, they are clearly linear time-invariant. Also,
pure delays are in this class since $e^{-\tau s} = \LL ( \mato{cc} 0 & 1 \\ 1 & 0 \matc , e^{-\tau s})$.

This GLTI class has two key properties, \cite{tdslft}:
\begin{itemize}
\item Any block diagram interconnection of GLTI systems is a GLTI system. In other words, the class of
GLTI systems is closed under series, parallel, and feedback connections as well as
branching/summing junctions.
\item The linearization of any nonlinear block diagram with time delays is a GLTI system.
\end{itemize}

These two properties show that the GLTI class is general enough to model any
(linearized) system with a finite number of delays, including delays in the
feedback path. For further motivation of this representation and equivalent case of discrete time systems, see \cite{tdslft}.

The GLTI class is represented in state-space equations as follows.
Let
\[
\left(
  \begin{array}{cc}
    H_{11}(s) & H_{12}(s) \\
    H_{21}(s) & H_{22}(s) \\
  \end{array}
\right)
  =
\left(
  \begin{array}{cc}
    D_{11} & D_{12} \\
    D_{21} & D_{22} \\
  \end{array}
\right)
  +
\left(
  \begin{array}{c}
    C_1 \\
    C_2 \\
  \end{array}
\right)
  (sI-A)\inv
\left(
  \begin{array}{cc}
    B_1 & B_2 \\
  \end{array}
\right)
\]
be a minimal realization of $H(s)$ in (\ref{sg:genlti_lft}).
State-space equations for $H(s,\tau) = \LL (H(s),\Theta(s,\tau))$ are readily
obtained as
\beqa
\left[
\begin{array}{c}
\dot{x}(t) \\
y(t) \\
z(t)
\end{array}
\right] &=&
\left[
\begin{array}{lll}
A & B_1 & B_2 \\
C_1 & D_{11} & D_{12} \\
C_2 & D_{21} & D_{22} \\
\end{array}
\right]
\left[
\begin{array}{c}
x(t) \\
u(t) \\
w(t)
\end{array}
\right] \label{sg:genlti1} \\
\nonumber w(t) &=& (\Delta_\tau z)(t)
\eeqa
where $u(t)$, $y(t)$ are the input and output vectors; $w(t)$, $z(t)$ are internal signals commensurate with the vector $\tau$ of time delays; $\Delta_\tau z$ is the vector-valued signal defined by
$(\Delta_\tau z) (t) := (z_1^T (t - \tau_1), \hdots, z_N^T (t - \tau_N))^T$.

Note that standard delay-free state-space models are just a special case of
(\ref{sg:genlti1}) corresponding to $N=0$, a handy fact when it
comes to integrating GLTI models with existing software for manipulating
delay-free state-space models.

Delay LTI systems of the form
\beqa
\nonumber \dot{x}(t) &=& A_0 x(t) + B_0 u(t) + \sum_{j=1}^M (A_j x(t-\theta_j) + B_j u(t-\theta_j))
\label{sg:lindde1} \\
\nonumber y(t) &=& C_0 x(t) + D_0 u(t) + \sum_{j=1}^M (C_j x(t-\theta_j) + D_j u(t-\theta_j))
\label{sg:lindde2}
\eeqa
are often considered in the literature with various restrictions on the number and locations of the
delays $\theta_1,\ldots,\theta_M$. It turns out that any model of this form belongs
to the class GLTI as shown in \cite{tdslft}. It is possible to define a large class of time delay systems in MATLAB, both in time and frequency domains. For further details on representation of time delay systems, see \cite{cst}.

\section{Interconnections} \label{sg:sec:intercon}
Control systems, in general, are built up by interconnecting other subsystems. The most typical configuration is a feedback loop with a plant and a controller as shown in Section~\ref{sg:sec:motivation}; whereas more complex configurations may have distributed systems with multiple plants, controllers and  transport / internal delays.

\begin{figure}[h]
\begin{center}
\resizebox{8.5cm}{!}{\includegraphics{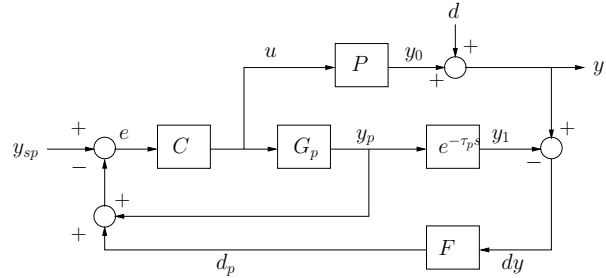}}
\caption{Feedback loop with Smith Predictor.\label{sg:fig:cs2}}
\end{center}
\end{figure}

A standard way to to obtain the closed-loop model of interconnections of systems in MATLAB is to use
the {\tt connect} command. This function requires all systems to have input and output names and
summation blocks. It automatically builds the resulting closed-loop system with the given
inputs and outputs. Consider the Smith Predictor control structure given in Figure \ref{sg:fig:cs2} for
the same dead-time system $P(s)$ in (\ref{sg:theplant}). The Smith Predictor uses
an internal model to predict the delay-free response $y_p(t)$ of the plant, and seeks to correct discrepancies
between this prediction and the setpoint $y_{sp}(t)$, rather than between the delayed output measurement
$y(t)$ and $y_{sp}(t)$. To prevent drifting, an additional compensator $F(s)$ is used to eliminate steady-state
drifts and disturbance-induced offsets.

We first assume that the prediction model $P_p(s) = e^{-\tau_p s} G_p(s)$ matches the plant model
$P(s)$ in (\ref{sg:theplant}), and use the following compensator settings:
\[
C(s) = 0.5 (1 + {1 \over 40 s}) , \;\;\; F(s) = {1 \over 20 s + 1}.
\]
By defining summation blocks and input and outputs names of systems, we obtain the closed-loop
model $T_{SP}$ from the input signal $y_{sp}$ to the output signal $y$:
\bg
\vspace{-.4cm}
{\small
\begin{verbatim}
  s = tf('s');

  % LTI blocks
  P = exp(-93.9*s) * 5.6/(40.2*s+1);
  P.InputName = 'u'; P.OutputName = 'y';

  Gp = 5.6/(40.2*s+1);
  Gp.InputName = 'u'; Gp.OutputName = 'yp';

  Dp = exp(-93.9*s);
  Dp.InputName = 'yp'; Dp.OutputName = 'y1';

  C = 0.5 * (1 + 1/(40*s));
  C.InputName = 'e';  C.OutputName = 'u';

  F = 1/(20*s+1);
  F.InputName = 'dy'; F.OutputName = 'dp';

  % Sum blocks
  Sum1 = ss([1,-1,-1],'InputName',...
         {'ysp','yp','dp'},'OutputName','e');
  Sum2 = ss([1,-1],...
         'InputN',{'y','y1'},'OutputN','dy');

  % Build interconnection model
  Tsp = connect(P,Gp,Dp,C,F,Sum1,Sum2,'ysp','y');
\end{verbatim}
}
\vspace{-.2cm}
\eg

We can also construct various types of connections such as in parallel and series ({\tt parallel} and {\tt series}); group systems by appending their inputs and outputs ({\tt append}); form the linear
fractional transformation ({\tt lft}). Standard system operations are also valid for time delay systems such as addition, subtraction, multiplication, division.

After we represent our subsystems and connect with each other, we easily construct the closed-loop model with time delays. Our next goal is to analyze the characteristics of the resulting closed-loop
models with visualizations and compute their system properties.

\section{Time / Frequency Domain Analyses and Visualizations} \label{sg:sec:analysis}

We analyze a plant or a closed-loop model with various interconnections and systems to understand its characteristics and properties. By simulating its time-domain response to certain inputs such as a step or tracking signals, we observe its time-domain characteristics such as rise and settling times, overshoot. On the other hand, frequency domain analysis gives us information on, for example, gain and phase margins, bandwidth and resonant peak.

In Section~\ref{sg:sec:representation}, we obtained the closed-loop system $T_{PI}$ of the dead-time system (\ref{sg:theplant}) and PI controller and in Section~\ref{sg:sec:intercon} we constructed the closed-loop system $T_{SP}$ of the same plant and the Smith Predictor. We simulate the responses of $T_{PI}$ and $T_{SP}$ to the tracking signal, {\tt ref} by the following commands:
\bg
\vspace{-.4cm}
{\small
\begin{verbatim}
% time and reference signal
time = 0:.1:2000;
ref = (time>=0 & time<1000)*4 + (time>=1000 & time<=2000)*8;

% compare responses
lsim(Tsp,Tpi,ref,time);
\end{verbatim}}
\vspace{-.2cm}
\eg

\begin{figure}[h]
 \begin{center}
 \resizebox{5.8cm}{!}{\includegraphics{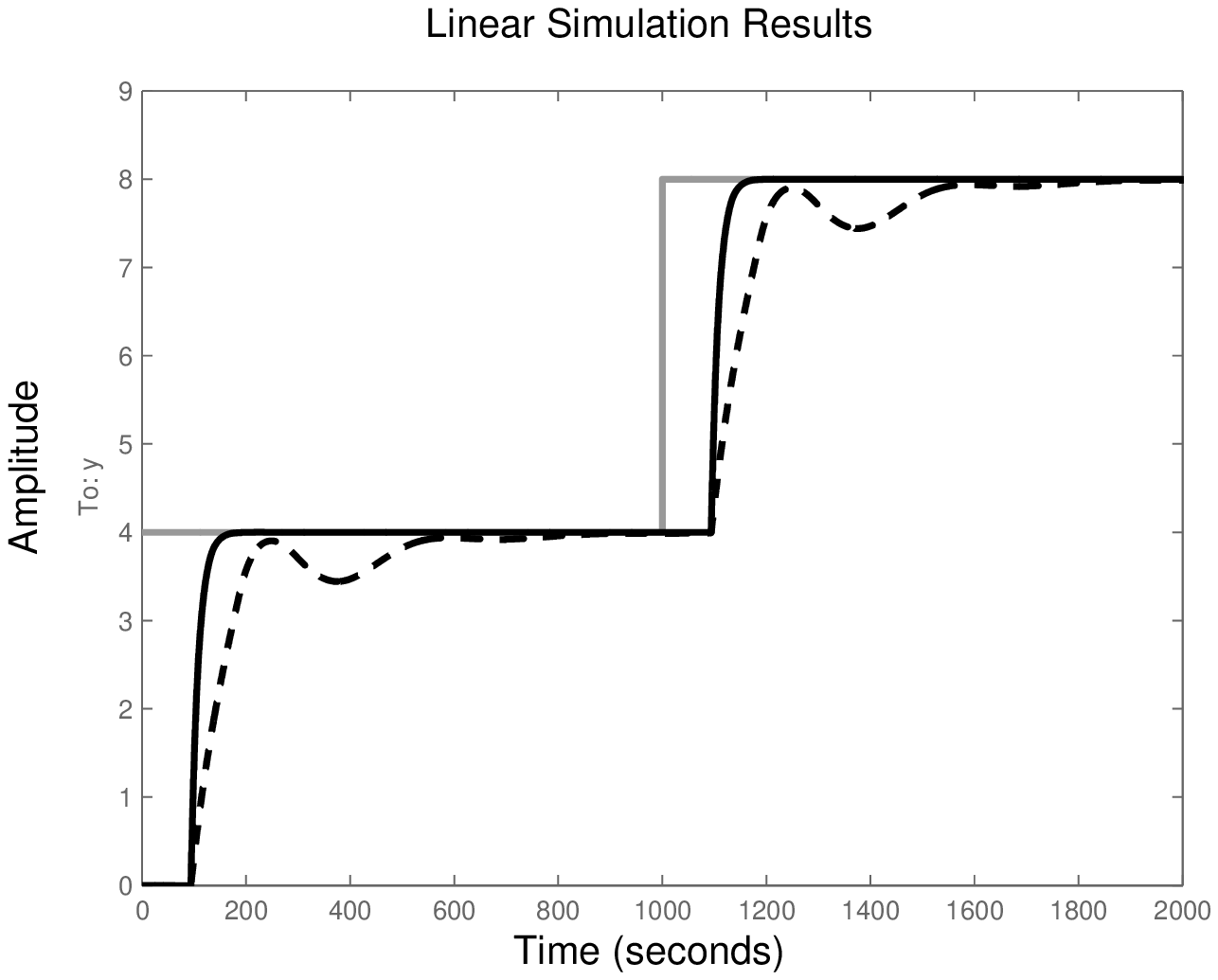}}
   \resizebox{5.8cm}{!}{\includegraphics{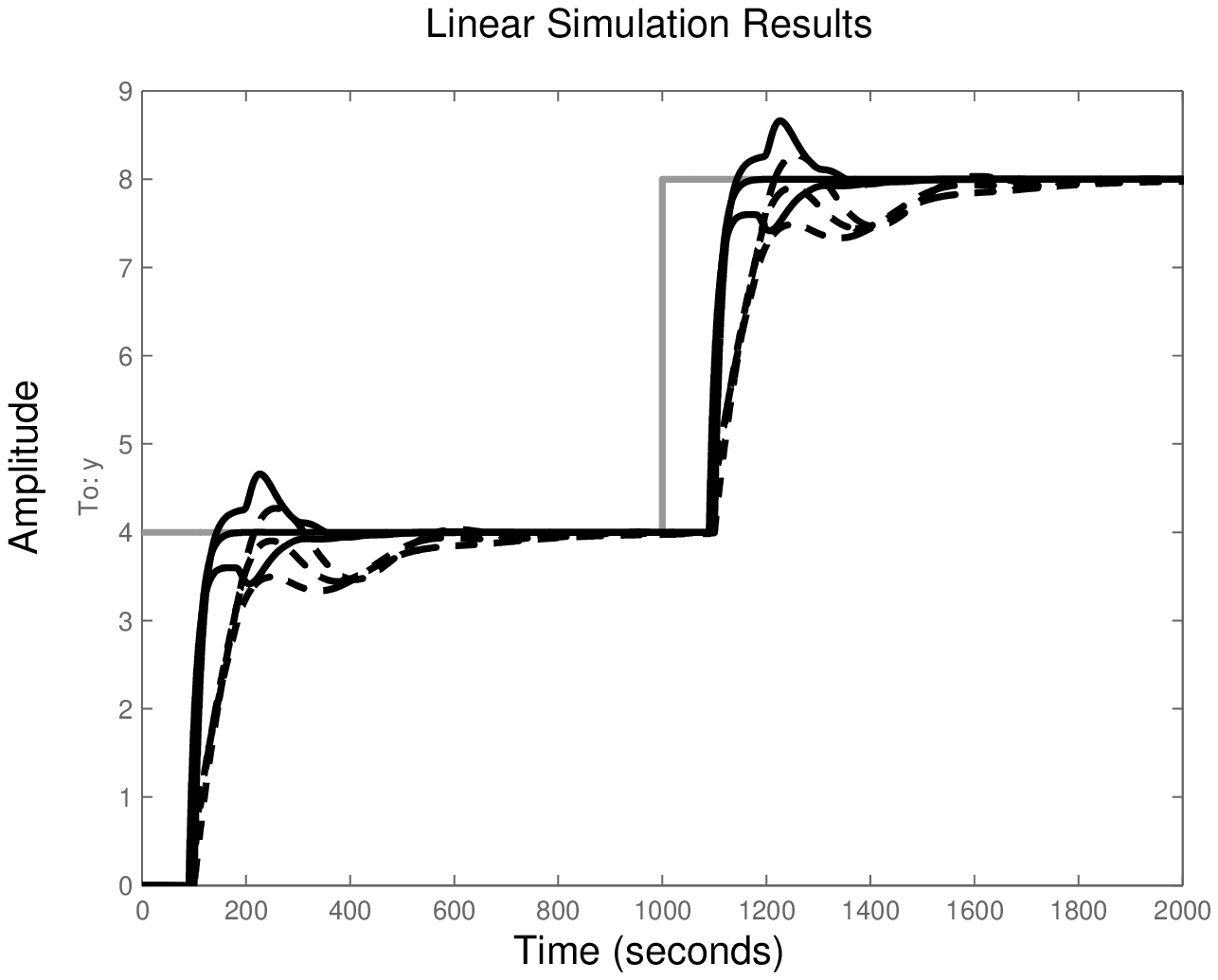}}
        \caption{ (left) Responses of the Smith Predictor (--) and PI (-\;-) to the reference signal (gray colored). (right) Robustness of the Smith Predictor (--) to Model Mismatch.         \label{sg:fig:trackresp}}
    \end{center}
\end{figure}

The resulting responses are shown in Figure \ref{sg:fig:trackresp} (on the left). Simulation results show that PI controller has a slower response time with oscillations and the Smith Predictor has better tracking performance.

In practice, there is always a mismatch between the predicted and real plant models. We easily investigate robustness of our design to modeling uncertainties. For example, consider two perturbed plant models
\[
P_1(s) = e^{- 90 s} {5 \over 38 s+1} , \;\;
P_2(s) = e^{- 100 s} {6 \over 42 s+1} \; .
\]
To assess the Smith predictor robustness when the true plant model is $P_1(s)$ or $P_2(s)$ rather than
the prediction model $P(s)$, simply bundle $P, P_1, P_2$ into an LTI array, rebuild the closed-loop model(s), and replot the responses for the tracking signal:
\bg
\vspace{-.4cm}
{\small
\begin{verbatim}
  P1 = exp(-90*s) * 5/(38*s+1);         % perturbed plants
  P2 = exp(-100*s) * 6/(42*s+1);
  Plants = stack(1,P,P1,P2);            % bundle true and perturbed plants

  T = connect(Plants,Gp,Dp,C,F,Sum1,Sum2,'ysp','y'); % construct closed-loop

  lsim(T,Tpi,ref,time);                 % simulate closed-loop responses
\end{verbatim}
}
\vspace{-.2cm}
\eg
\noindent
The resulting responses in Figure \ref{sg:fig:trackresp} (on the right) show a slight performance degradation, but the Smith predictor
still retains an edge over the pure PI design.

\begin{figure}[h]
\centerline{\includegraphics[scale=.5]{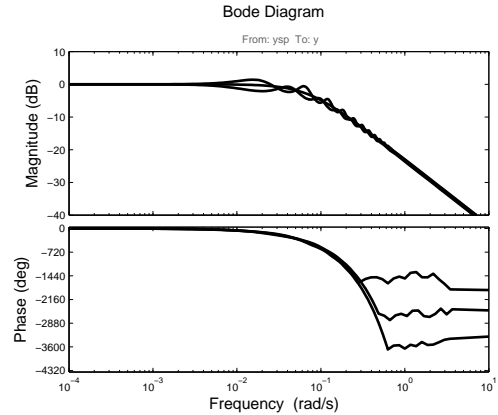}}
\caption{Closed-Loop Response from $y_{sp}$ to $y$. \label{sg:fig:bode}}
\end{figure}

We obtain the closed-loop frequency responses for the nominal and perturbed plants by {\tt bode(T)} and their visualizations as shown in Figure \ref{sg:fig:bode}. Note that the phase behavior of systems with internal delays is quite different than systems with I / O delays.

We numerically compute the bandwidth of the responses by {\tt bandwidth(T)} which
returns $0.0695$, $0.0565$, $0.0767$. The gain and phase margins of the responses are
calculated by {\tt [gm, pm] = margin(T)} and their values are
\[
{\tt gm} = [1.0835;\ \ 1.1569;\ \ 1.0304],\quad
{\tt pm} = [180;\ \ 180;\ \ 7.1433].
\] Other well-known frequency-domain based tools are also available for the GLTI class such as {\tt bandwidth}, {\tt dcgain}, {\tt nyquist}, {\tt allmargin}.

\section{Controller Design} \label{sg:sec:controller}

We analyzed the closed-loop characteristics of the given PI controller and Smith Predictor. Now we design a PID controller for finite dimensional and time delay plants using {\tt pidtune} function in The Control System Toolbox and compare its performance with other controllers. This function aims to find a PID controller stabilizing the closed-loop system and to satisfy certain performance and robustness objectives. These objectives are tracking reference changes and suppressing disturbances as rapidly as possible; designing enough phase and margins for modeling errors or variations in system dynamics.

The algorithm for tuning PID controllers helps us meet these objectives by automatically tuning the PID gains to balance the response time as a performance objective and the stability margins as robustness objectives. By default, the algorithm chooses a crossover frequency (loop bandwidth) based upon the plant dynamics, and designs for a target phase margin of $60^\circ$.

\begin{figure}[h]
 \begin{center}
 \resizebox{5.8cm}{!}{\includegraphics{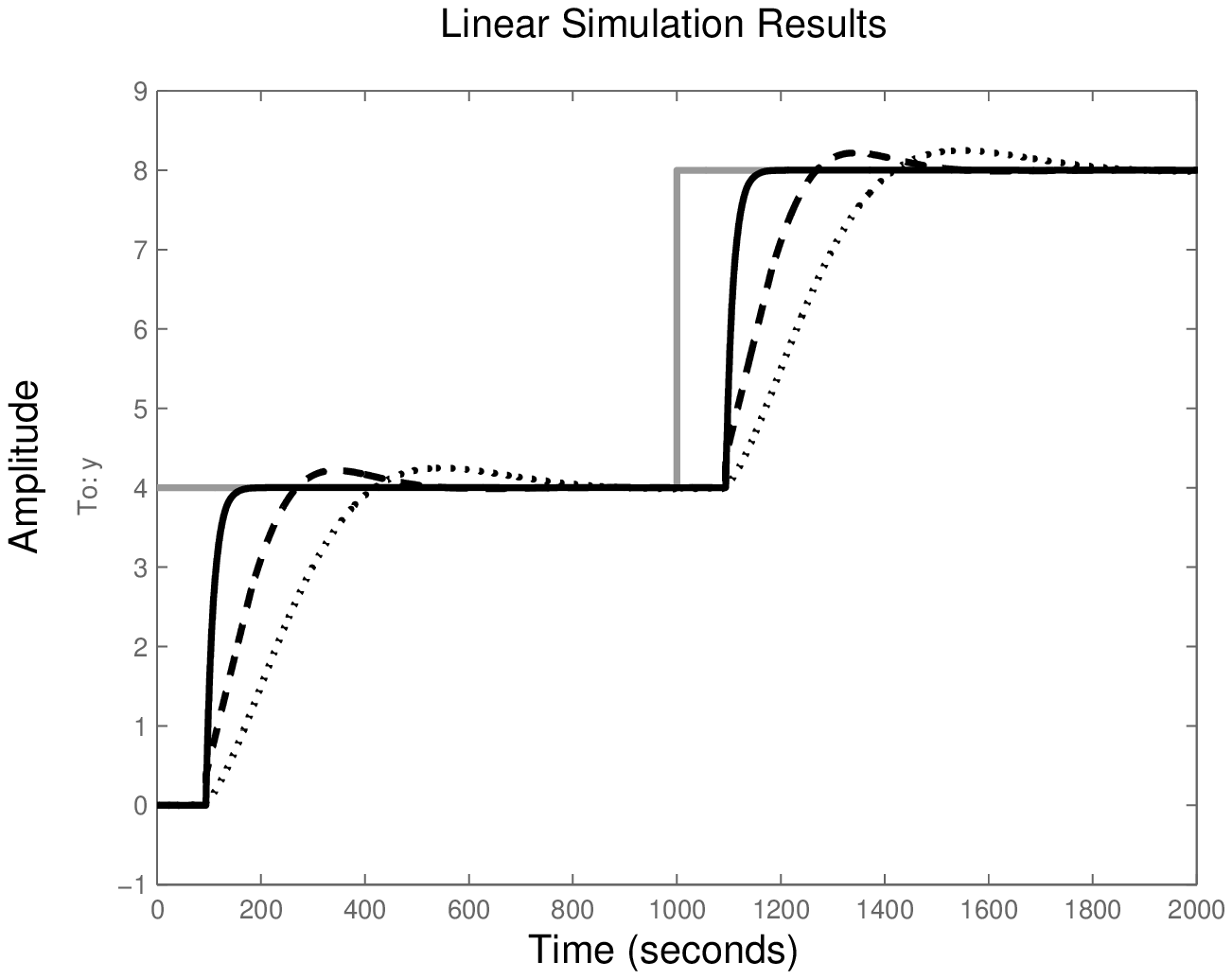}}
   \resizebox{5.8cm}{!}{\includegraphics{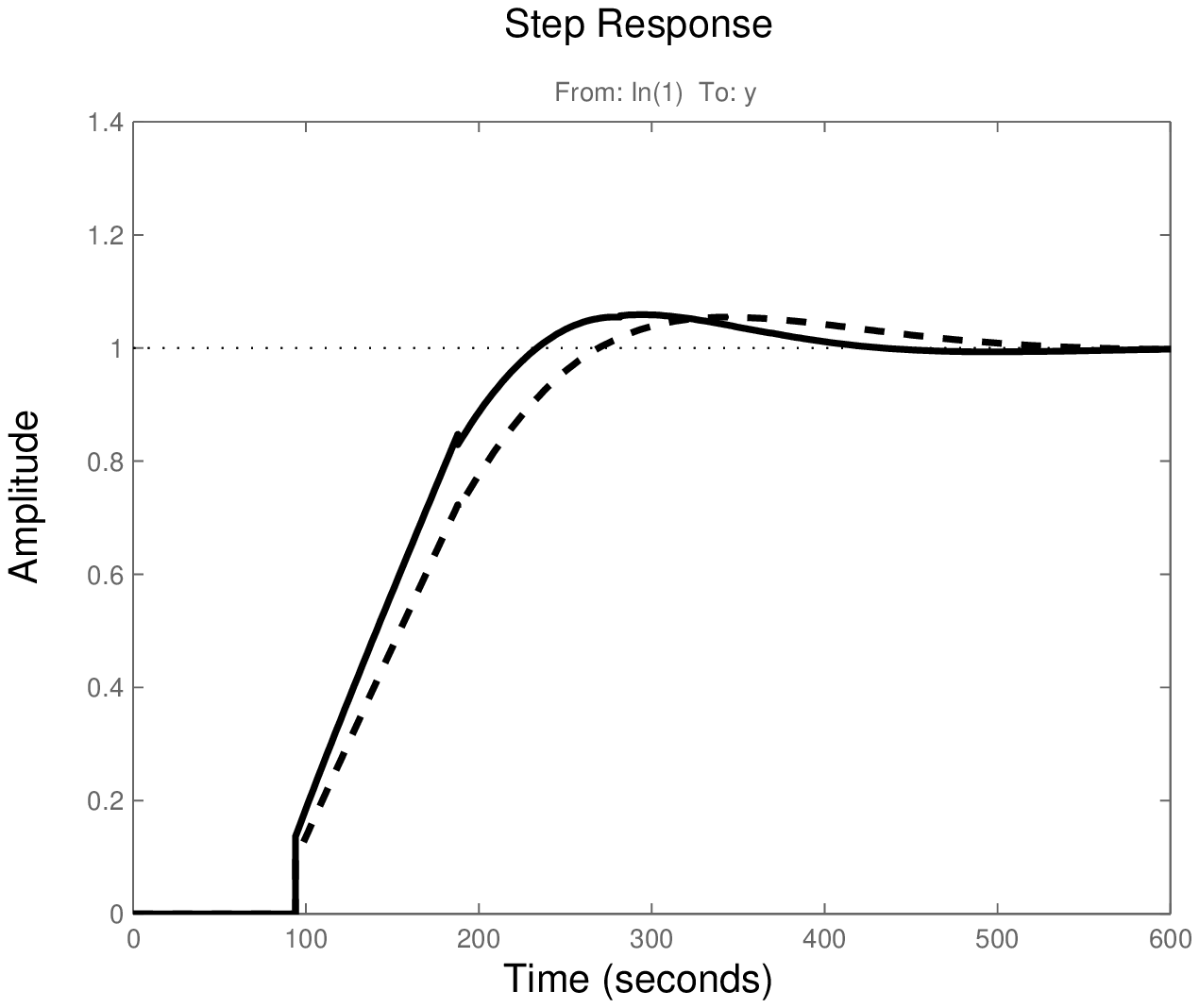}}
        \caption{ (left) Responses of the Smith Predictor (--), PID (-\;-) for $P(s)$ and PID (:) for $P_a(s)$ to the reference signal (gray colored). (right) Closed-loop step responses for the controllers  {\tt Cpid} (--) and {\tt Cpidf} (-\;-).         \label{sg:fig:stepresp}}
 \end{center}
\end{figure}

We can approximate the dead-time system $P(s)$ by a finite dimensional transfer function $P_a(s)$ using the function {\tt pade} based on Pad{\'{e}} approximation. The function {\tt pidtune} designs a PID controller $C_a(s)$ for the approximate finite dimensional plant and we obtain the closed-loop system for this controller by the following commands:
\bg
\vspace{-.4cm}
{\small
\begin{verbatim}
  Pa = pade(P,8);         % approximate 8th order plant
  Ca = pidtune(Pa,'pid'); % design PID for Pa
  Ta = feedback(P*Ca,1);  % closed-loop for Ca
\end{verbatim}}
\vspace{-.2cm}
\eg
{\tt Pidtune} also designs a PID controller for time delay systems \emph{without any approximation},
\bg
\vspace{-.4cm}
{\small
\begin{verbatim}
  [Cpid,info] = pidtune(P,'pid');   % design PID for P
  Tpid = feedback(P*Cpid,1);        % closed-loop for Cpid
  >> info

  info =

                  Stable: 1
      CrossoverFrequency: 0.0067
             PhaseMargin: 60.0000
\end{verbatim}
} 
\vspace{-.2cm}
\eg
As shown in returned {\tt info} structure, the designed controller {\tt Cpid} stabilizes the closed-loop and achieves $0.0067$ rad/sec crossover frequency and $60^\circ$ phase margin. The closed-loop step response {\tt Tpid} of the controller {\tt Cpid} is given Figure~\ref{sg:fig:stepresp} (on the right with dashed line). Through step plot figure, we compute its transient response characteristics. The step response for this controller has $5.45\%$ overshoot, $136$ and $459$ seconds rise and settling times.

We compare the closed-loop responses of the Smith Predictor, the designed PID controllers for the approximate plant $P_a(s)$ and the original plant $P(s)$ by
\bg
\vspace{-.4cm}
{\small
\begin{verbatim}
  lsim(Tsp,Ta,Tpid,ref,time);
\end{verbatim}
} 
\vspace{-.2cm}
\eg The responses in Figure~\ref{sg:fig:stepresp} (on the left) show that the designed PID controller for the original plant offers a good compromise between the simplicity of the controller and good tracking performance compared to the Smith predictor.

We can fine tune the PID controller depending on design requirements. If faster response is required, we can increase the crossover frequency slightly and obtain the controller {\tt Cpidf} by
\bg
\vspace{-.4cm}
{\small
\begin{verbatim}
Cpidf = pidtune(P,'pid',0.0074);
Tpidf = feedback(P*Cpidf,1);
step(Tpid,Tpidf);
\end{verbatim}
} 
\vspace{-.2cm}
\eg
The closed-loop step responses for the controllers  {\tt Cpid} and {\tt Cpidf} are shown Figure~\ref{sg:fig:stepresp} (on the right). The new controller {\tt Cpidf} has a faster response where its rise and settling times are $109$ and $394$ seconds, almost $20\%$ and $17\%$ faster than that of {\tt Cpid} and its overshoot is slightly increased from $5.45\%$ to $5.89\%$.

\section{Possible Enhancements in CACSD} \label{sg:sec:plan}

We briefly summarized the available functionality in Control System Toolbox and we discuss possible enhancements for CACSD regarding time delay systems in this section. As illustrated before, most of the functions in Control System Toolbox are extended for time delay systems. We focus on three important numerical computations for time delay systems and discuss on-going research directions on these computations.

\subsubsection*{Stability of a time delay system}

There are various numerical methods to determine the stability of LTI systems with constant delays \cite{Dugard1998,GuBook,WimBook}. One idea is to compute the characteristic roots when time delay is zero and to detect characteristic roots crossing the imaginary axis from zero delay to desired time delay and determine the stability of the time delay system. This approach is applicable to only systems with commensurate time delays and quasi-polynomial form is required. Another approach is to approximate the right-most characteristic roots in the complex plane using spectral methods further explained in the next section. The computational cost in this method depends on the number of discretization points for the time delay interval, i.e., from zero to maximum delay in the system. There are some heuristic methods to choose this number and they may result in poor choices at certain cases. Lyapunov theory is another tool to determine the stability of time delay systems. The results are conservative and in general the conservatism can be reduced in the expense of the computational cost of solving larger linear matrix inequalities. Most methods in the literature can not handle time delay systems with high orders.

\subsubsection*{System poles and zeros}

The poles and zeros of time delay systems are computed by solving a nonlinear eigenvalue problem, essentially same problem to compute the characteristic roots of time delay systems. Therefore, approximating spectrum approach for characteristic roots is also used to compute system poles and zeros.

The computations are based on either discretization of the solution operator of a delay differential equation or the infinitesimal generator of the solution operator semigroup. The solution operator approach by linear-multi-step time integration for retarded type delay differential equations is given in \cite{Engelborghs2002,BIFTOOLManual20}. The infinitesimal generator approach discretizes the derivative in abstract delay differential equation by Runge-Kutta or linear multi-step methods and approximates into a matrix \cite{BredaSISC2005,BredaTraceDDE09} for retarded type delay differential equations with multiple discrete and distributed delays. Extensions to neutral type delay differential equations and mixed-type functional differential equations are done in \cite{BredaANM06}. Numerically stable implementation of spectral methods with some heuristics is given in \cite{TW596}.

The computation of system poles and zeros is closely connected with the nonlinear eigenvalue problem and an eigenvalue algorithm for this is presented in \cite{TW580}. A numerical method to compute all characteristic roots of a retarded or neutral quasi-polynomial on a large region in the complex-plane is proposed in \cite{TomasQPmR}. The characteristic roots are calculated by finding the intersection of real and imaginary part of the characteristic equation on certain regions in complex-plane. This approach is further improved and accelerated by removing the regions outside of asymptotic chain roots in \cite{VZTAC09}. These methods consider the transfer function representation of delay differential equations which can be written as a ratio of quasi-polynomials. As noted in \cite{VZTAC09}, when delay differential equations have state-space representations, transforming these systems into transfer function representation is not numerically desired, therefore in this case discretization approaches may be preferred.

\subsubsection*{$\Hi$ and $\mathcal{H}_2$ \emph{norms}}

The computation of $\Hi$ and $\mathcal{H}_2$ norms of time delay systems are quite new research topics and there are few research papers on these topics. Similar to the computation of system poles and zeros, $\Hi$ norm computation is reduced solving a nonlinear eigenvalue problem where the recent developments are applicable, \cite{Michiels2010}. The computation of $\mathcal{H}_2$ norm requires solving the delay Lyapunov equation, \cite{JarlebringTAC11}.

Note that all three computation methods are mainly used for analysis of time delay systems. Another challenging task is to design controllers and to extend classical control methods to time delay systems such as LQG, $\mathcal{H}_2$ control, $\Hi$ control, root-locus technique, model reduction methods.

There are continuing research efforts to solve these problems such as \cite{Vanbiervliet2011IJC,Gumussoy2011SICON,Gumussoy2012AUT,TW602}. The remaining main task is to determine numerically stable algorithms to solve control design problems for high dimensional plant with the least user interactions.

\section{Concluding Remarks} \label{sg:sec:conclusion}

We have shown that the GLTI class is suitable for computer-aided manipulation of time delay systems. We discussed various representations and interconnections of time delay systems on MATLAB. We presented the MATLAB functionality to analyze and design control systems with delays, regardless of the control
structure and number of delays. Most Control System Toolbox functions have been extended to work on GLTI models, all this without additional complexity or new syntax for the user. We hope that these new tools will facilitate the design of control systems with delays and bring new insights into their behavior.

%%%  Do not use Bibtex for your citations
%%%  please use the following:
%SEE referenc.tex and authsamp.pdf FILES FOR EXAMPLES OF CITATION STYLES

\end{document}